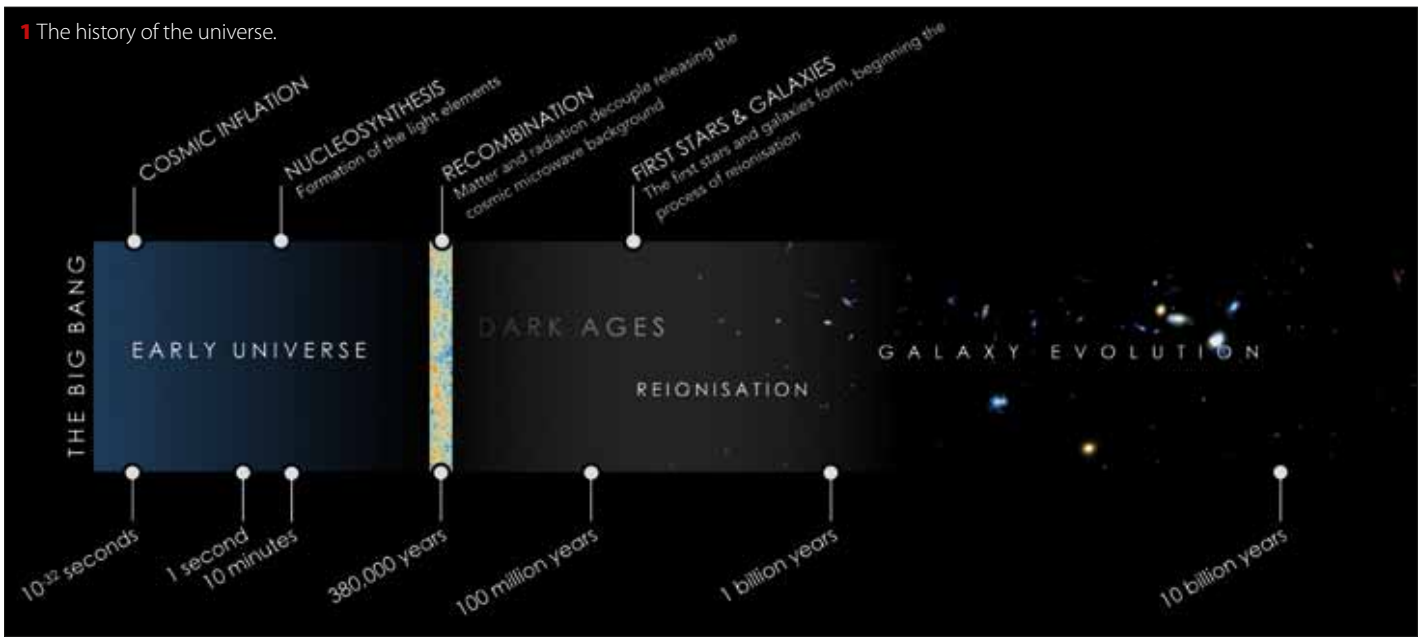

1 The history of the universe.

# Exploring the dawn of galaxies

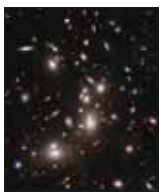

**Stephen M Wilkins** and **Elizabeth Stanway** set the scene for the themed set of articles in this issue, looking back to the cosmic dawn, just a few hundred million years after the Big Bang, when galaxies first formed.

Some few hundred million years after the Big Bang, the universe was illuminated by the first stars and galaxies, thereby bringing an end to the cosmological dark ages. Since the installation of the Wide Field Camera 3 on the Hubble Space Telescope, our ability to probe this critical period of the universe's history has changed dramatically, with thousands of objects now identified within the first billion years of the universe's history. Both the Atacama Large Millimetre/submillimetre Array and the James Webb Space Telescope will add to the observational data. Researchers can now investigate this cosmic dawn.

The current consensus in cosmology is that our universe emerged from an initially hot, dense, plasma state roughly 14 billion years ago. After some 380 000 years the universe had expanded and thus cooled sufficiently to allow neutral hydrogen to form (the process of recombination). Shortly thereafter, photons decoupled from matter and began to travel virtually unimpeded through the universe. Photons streaming from this "surface of last scattering" are observed today as the cosmic microwave background, providing a snapshot of the early universe (figure 1).

The formation of hydrogen marks the beginning of the cosmological dark ages. During this time the only source of electromagnetic radiation was the 21 cm fine structure line of neutral hydrogen, the observation of which is a primary driver of the next generation of radio telescopes (including the Square Kilometre Array and its pathfinders, see article by Jonathan Pritchard on pages 3.25–3.30). These are expected to become a powerful tool for observational cosmology (Pritchard *et al.* 2015).

During the dark ages, structures developed through collapse of matter onto gravitational instabilities. Eventually the dark ages were brought to an end as the first stars, galaxies, and active galactic nuclei (AGN) began to illuminate their surroundings some few hundred million years after recombination (see article by Iliev, Sullivan and Dixon on pages 3.31–3.33). The intense radiation produced by these sources eventually (by redshift $z \sim 6$–$7$, 1 billion years after the Big Bang) reionized the universe, leaving its baryonic component once again dominated by ionized hydrogen (see Bouwens *et al.* 2015 for a recent observational overview).

Identifying early galaxies, and confirming them as the source of the photons responsible for reionization, has long been a major goal of observational extragalactic astronomy.

> "Photons from the surface of last scattering provide a snapshot of the early universe"

## The Lyman-break technique

The Lyman-break technique is popular for identifying UV-bright high-redshift star-forming galaxies. The basic technique relies on the fact that electromagnetic radiation at higher energies than the Lyman limit (91.2 nm) is efficiently scattered by neutral gas surrounding star-forming regions of galaxies. This results in a strong break at 91.2 nm in the rest frame spectrum. A star-forming galaxy visible in images obtained using a filter longward of the break would effectively "drop-out" in shorter wavelength observations. At very high redshifts, regions of neutral hydrogen in the intergalactic medium also suppress electromagnetic radiation with wavelengths less than the Lyman-$\alpha$ transition (121.6 nm). For a galaxy at $z \sim 7$ the Lyman-$\alpha$ limit falls approximately between the z and Y-bands and $z \sim 7$ galaxies are sometimes called z-dropouts (see example in the figure).

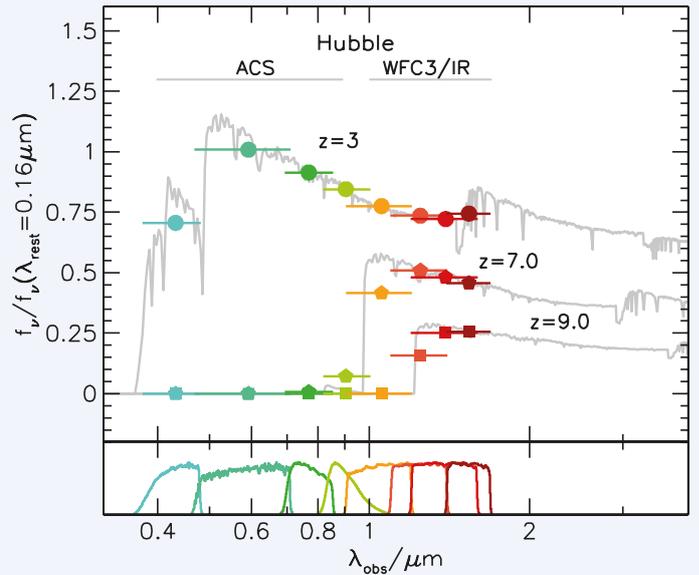

**2** The observed frame spectral energy distribution and photometry expected for a star-forming galaxy at $z = 3$, 7 and 9. Galaxies at $z \sim 7$ are likely to be undetectable at optical wavelengths.

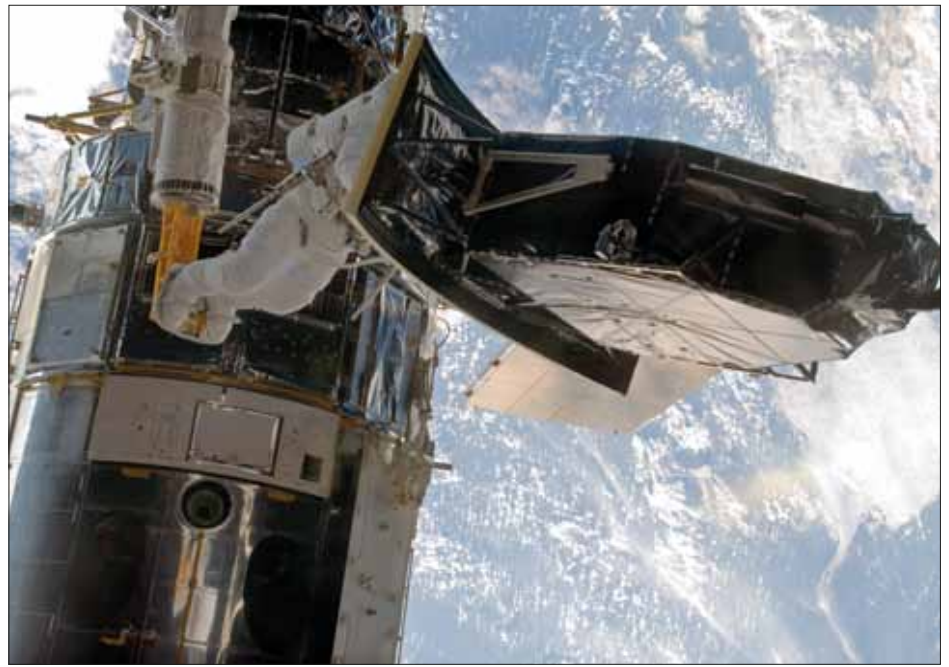

**3** The astronaut Andrew Feustel helps to install the Wide Field Camera 3 (WFC3) on the Hubble Space Telescope (HST). WFC3 is one of the most advanced instruments on the HST. It was installed during the fifth and final space shuttle servicing mission to HST in May 2009. WFC3 offers the ability to obtain imaging and low-resolution grism spectroscopy over a very wide wavelength range with a large field of view. The UV/visible (UVIS) channel of WFC3 covers both the visible spectrum (380 nm to 780 nm) and the near-ultraviolet (down to 200 nm) with a field of view of 2.7 × 2.7 arcmin. The near-infrared channel has a field of view of 2.3 × 2.1 arcmin and extends to 1700 nm.

### Finding galaxies in the early universe

The discovery and study of galaxies during the epoch of reionization ($z \sim 7$) is made incredibly challenging by the distances involved. The luminosity distance to $z = 7$ is ~230 billion light years; even intrinsically luminous objects appear incredibly faint. For this reason the high-redshift field has always pushed the frontiers of instrument development.

One of the first major milestones in the study of galaxies in the early universe was the identification, by Steidel *et al.* (1996), of a population of galaxies at $z \sim 3$ (approximately 2 billion years after the Big Bang) using the now ubiquitous Lyman-break technique (see box). Subsequent surveys built upon this pioneering early work by expanding sample sizes and pushing to higher redshifts.

Despite the tremendous progress in those early years it was not until the installation of the Advanced Camera for Surveys (ACS) on the Hubble Space Telescope in 2002 that significant progress was made towards the predicted epoch of reionization. Hubble/ACS allowed astronomers to obtain deep, wide-area imaging in the near-infrared (near-IR) z-band (~850 nm), allowing the efficient selection of galaxies at $z \sim 5$–6 (e.g. Stanway *et al.* 2003, Bouwens *et al.* 2003, Bremer *et al.* 2004) within the first billion years of the history of the universe.

Subsequent progress in pushing to $z \sim 7$ and beyond was slowed by the poor survey efficiencies of near-IR cameras on both ground-based telescopes and Hubble, with only a handful of $z \sim 7$ galaxy candidates identified prior to 2009. This situation changed dramatically with the installation of the near-IR sensitive Wide Field Camera 3 (WFC3, figure 3) on the HST in 2009. The superb sensitivity, field of view and spatial resolution of this camera dramatically improved the efficiency with which Hubble could survey the sky in the near-IR (e.g. Oesch *et al.* 2010, Bouwens *et al.* 2010, Bunker *et al.* 2010, Wilkins *et al.* 2010, 2011).

Since the installation of WFC3, a number of programmes have been undertaken whose science goals have focused on identifying and characterizing galaxies at $z \sim 7$ and above. Most prominent among these,

> "The deepest VISTA surveys have identified a sample of bright $z \sim 7$ galaxies"

**4** The Atacama Large Millimetre/submillimetre Array (ALMA) is an interferometric radio telescope in which the data from many dishes is combined to increase the effective resolution over that of a single dish. It is at an advanced stage of commissioning, sited in the Atacama Desert in northern Chile. ALMA is an international initiative consisting of partners from Europe, North America, East Asia and the Republic of Chile. (ESO/B Tafreshi http://twanight.org)

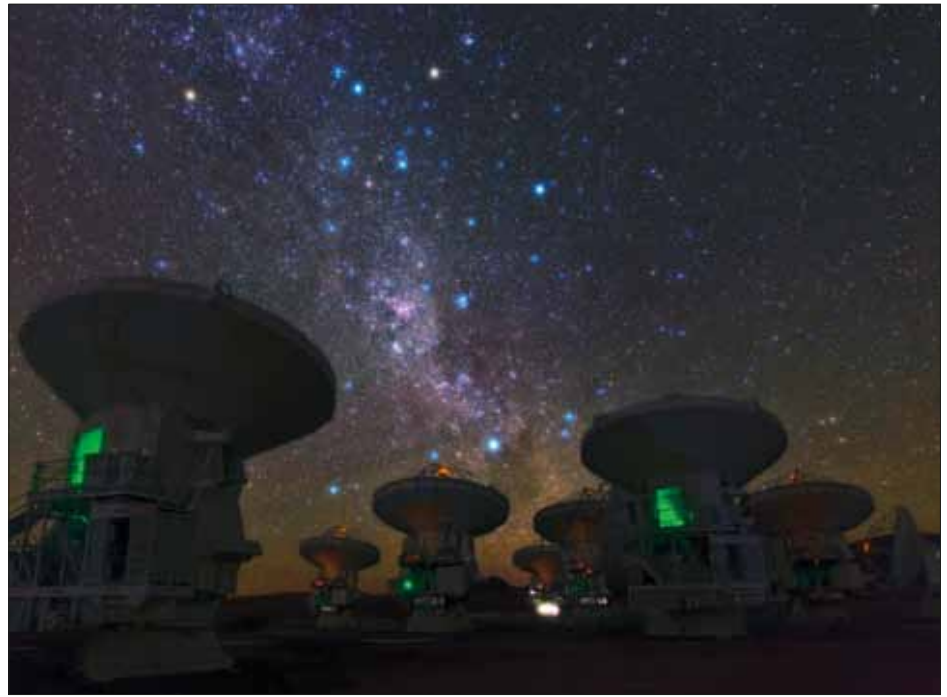

in terms of their effect on the discovery of distant galaxies, have been WFC3/IR observations of the Hubble Ultra Deep Field (HUDF) and the Cosmic Assembly Near-infrared Deep Extragalactic Legacy Survey (CANDELS). Combined, these programmes have now uncovered several hundred candidate galaxies at $z \sim 7$ and $z \sim 8$ (e.g. Bouwens *et al.* 2015) with the first, albeit small, samples proposed at $z \sim 10$ (e.g. Oesch *et al.* 2013) – approximately 500 million years after the Big Bang.

Large Hubble/WFC3 campaigns targeting the early universe continue most prominently under the Frontier Fields programme, which targets the fields of known, low-redshift galaxy clusters and parallel blank fields. The cluster fields offer the exciting opportunity to exploit gravitational lensing, which magnifies and enables the study of intrinsically faint sources (see article by Renske Smit on pages 3.34–3.38).

While Hubble/WFC3 sheds new light on the high-redshift universe, its ability to discover the critical population of rare, bright sources is hampered by its field of view. Such sources are highly valued as they are amenable to multi-wavelength follow-up in detail or with less sensitive facilities. Ground-based facilities, in particular the combination of the Visible and Infrared Survey Telescope for Astronomy (VISTA) and its sensitive near-IR camera, are working to fill this gap. Using observations from the deepest VISTA survey (UltraVISTA), a sample of bright $z \sim 7$ galaxies has been identified (see article by Rebecca Bowler on pages 3.39–3.43, also Bowler *et al.* 2015), significantly extending the luminosity baseline of identified high-redshift galaxy candidates.

**5** The predicted observed optical–millimetre spectral energy distribution of a high-redshift star-forming galaxy, highlighting the regions accessible to Hubble, Spitzer and ALMA.

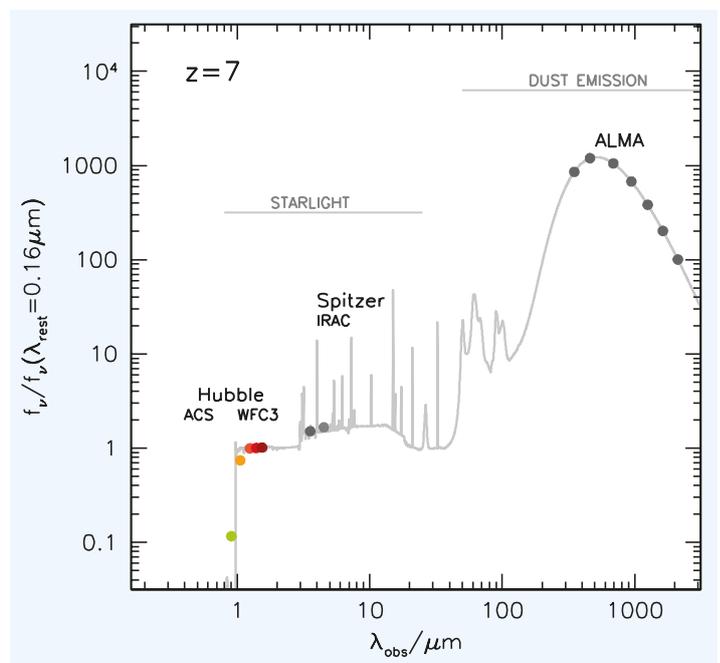

### Spectroscopic confirmation

Because of the possibility of contamination from lower redshift or galactic interlopers (e.g. Wilkins *et al.* 2014), spectroscopic confirmation of at least a representative sample of the $z \sim 7$ and above sources has been a key priority. This is most efficiently achieved by identifying a bright emission line, since the spectral continuum is often too faint. Unfortunately, only one strong hydrogen transition (Lyman-$\alpha$, rest-frame 121.6 nm) is currently accessible at $z \sim 7$ and has proven elusive in many of the campaigns pursuing samples of candidate high-redshift galaxies (see article by Joseph Caruana on pages 3.44–3.46).

### Star formation in the early universe

The star-formation activity (or rate) is a key physical property of galaxies (and indeed the entire universe) and is naturally predicted by galaxy formation models. Hubble and VISTA observations of galaxies at $z \sim 7$ and beyond probe only the rest-frame UV emission of galaxies and, while the intrinsic UV is an established tracer of star formation, it is severely attenuated by dust. The observed rest-frame UV luminosities of galaxies thus provide only a lower limit on the intrinsic star-formation activity. The degree of dust attenuation may be estimated from rest-frame UV continuum slope observations, but this technique is dependent on a range of assumptions limiting its usefulness, particularly in the early universe (see, for example, Wilkins *et al.* 2013).

More robust constraints come from combining rest-frame UV and far-infrared photometry (the far-IR probes emission reprocessed by dust). Until recently obtaining far-IR photometry of any but the most luminous systems at high-redshift was impossible. This has changed with the commissioning of the Atacama Large Millimetre/submillimetre Array (ALMA, figure 4). Submillimetre observations benefit from a "negative k-correction" – the balanced effects of increasing luminosity distance against observations at a frequency closer to the peak of a thermal dust emission curve – and so ALMA provides the sensitivity and wavelength coverage (figure 5) to directly probe the far-IR emission of high-redshift galaxies (see Watson *et al.* 2015 for an impressive recent detection at $z>7.5$).

By measuring the intrinsic star-formation rates of a representative sample of galaxies, it is possible to constrain the star-formation rate density (i.e. the average amount of star formation per unit volume). Doing so at multiple epochs then allows us to map out the history of star formation in the universe (see Madau & Dickinson 2014 for a recent review). This observed cosmic star-formation history (see figure 6) shows an initially rapid increase to $z \sim 2$–$3$ before flattening and then declining towards the present day.

### The future

Hubble and VISTA will, over the coming years, continue to identify galaxies in the distant universe. However, both are effectively limited to observing objects at $z<11$, beyond which the emission of these galaxies is redshifted beyond the current instrumental capabilities.

Overcoming this barrier requires the ability to obtain sensitive observations at wavelengths >1700 nm. Such capability will, in the near future, be provided by the James Webb Space Telescope (JWST, figure 7), due to be launched in 2018. JWST is expected to identify thousands of galaxies at $z>7$ including the first samples at $z \sim 12$ and potentially beyond. JWST will also provide exquisite multi-wavelength observations including deep near- and mid-IR imaging.

JWST's study of the high-redshift universe will be complemented by ALMA's work at submillimetre wavelengths and a range of other upcoming facilities such as the new generation of Extremely Large Telescopes (e.g. the European ELT). These next-generation ground-based facilities will provide exquisite spectroscopy of even the most distant galaxies. ●

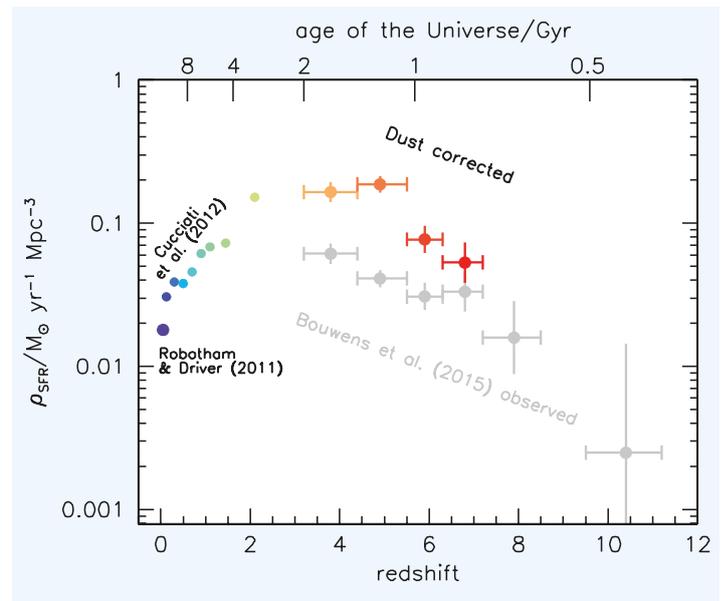

**6** The evolution of the star formation rate density as probed by the dust-corrected UV luminosity density.

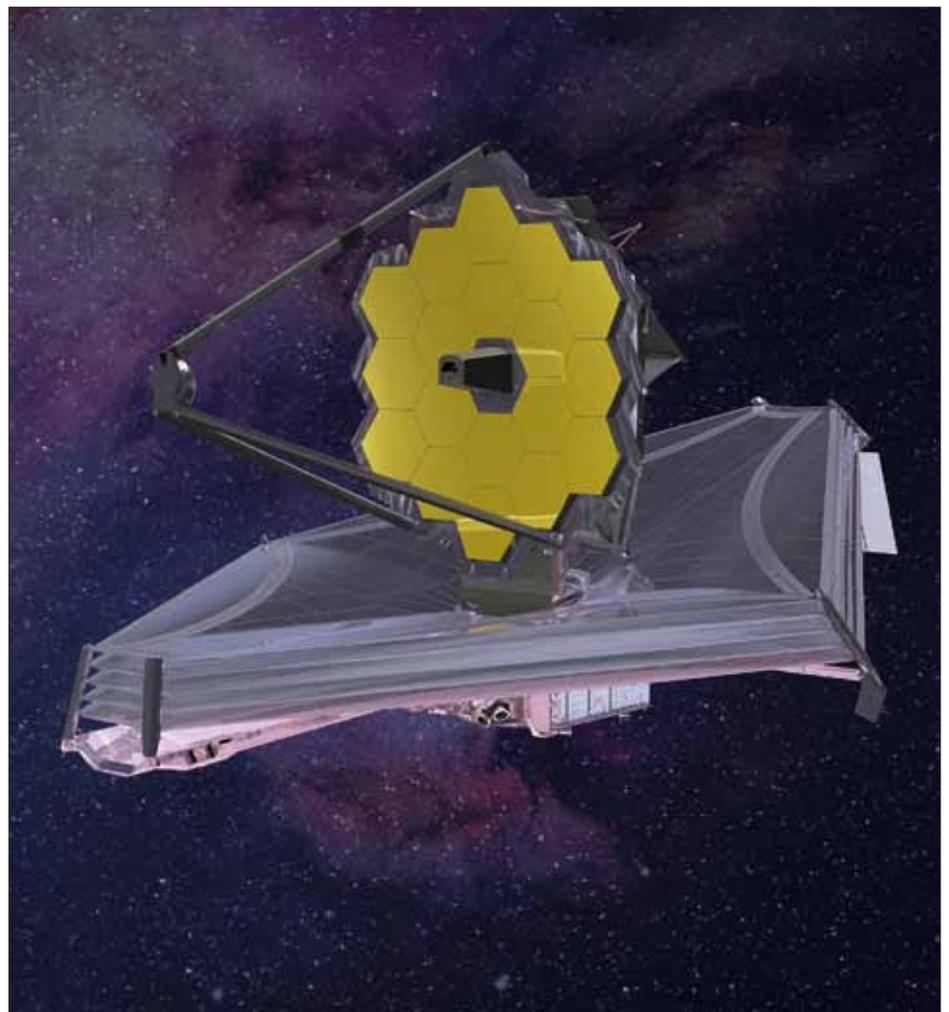

**7** The James Webb Space Telescope (JWST) is a space observatory with a planned launch date of late 2018. JWST will, through a combination of instruments, offer unprecedented resolution and sensitivity from the long-wavelength optical to the mid-infrared and is the successor to both the Hubble Space Telescope and Spitzer Space Telescope. JWST features a segmented 6.5 m primary mirror, significantly larger than either of its predecessors (Hubble: 2.5 m; Spitzer: 0.85 m). (Northrop Grumman)

**AUTHORS**
**Stephen M Wilkins**, Astronomy Centre, Department of Physics and Astronomy, University of Sussex, UK.
**Elizabeth Stanway**, Department of Physics, University of Warwick, UK.